\documentstyle{article}
\newtheorem{rem}{Remark}
\newtheorem{theor}{Theorem}

\begin{document}
\begin{center}
\Large{\bf FOUR DIMENSIONAL EINSTEIN SPACES ON \\[2mm] SIX
DIMENSIONAL RICCI FLAT BASE SPACES }\vspace{4mm}\normalsize
\end{center}
 \begin{center}
\Large{\bf Valery Dryuma}\vspace{4mm}\normalsize
\end{center}
\begin{center}
{\bf Institute of Mathematics and Informatics AS Moldova, Kishinev}\vspace{4mm}\normalsize
\end{center}
\begin{center}
{\bf  E-mail: valery@dryuma.com;\quad cainar@mail.md}\vspace{4mm}\normalsize
\end{center}
\begin{center}
{\bf  Abstract}\vspace{4mm}\normalsize
\end{center}

   Some examples of ten-dimensional vacuum Einstein spaces ($^{10}R_{ij}=0$)
   composed of four-dimensional Ricci-flat ($^{4}R_{ij}=0$) Einstein spaces and
    six-dimensional Ricci-flat  base spaces ($^{6}R_{ij}=0$) defined by the solutions of the classical
Korteveg-de Vries equation are constructed.

    The properties of  geodesics of such type of the spaces are studied.

\section{Introduction}

   The properties of classical four dimensional  Einstein spaces are determined by the
   energy-momentum tensor of  matter $T_{ik}$
\begin{equation} \label{dryuma:eq1}
R_{i j}=\frac{8
\pi\kappa}{c^4}\left(T_{ik}-\frac{1}{2}g_{ik}T\right).
\end{equation}

    Tensor $T_{ik}$ is the self-dependent object in the Einstein theory of gravitation and in general
    does not has geometric description.

    The most popular approach to the geometric description of the
    tensor $T_{ik}$ takes place within the bounds of the Kaluza-Klein theories using
     the string theory on the Calaby-Yau manifolds as the component.

    We present here a new possibilities to such type of considerations.

\section{Three-dimensional space of zero curvature}

   We start from three dimensional space with the metrics in form
\begin{equation} \label{dryuma:eq2}
{{\it ds}}^{2}={y}^{2}{{\it dx}}^{2}+2\,\left
(l(x,z){y}^{2}+m(x,z) \right ){\it dx}\,{\it dz}+2\,{\it dy}\,{\it
dz}+\]\[+\left (\left (l(x,z) \right )^{2}{y}^{2}-2\,\left ({\frac
{\partial }{\partial x}}l(x,z) \right
)y+2\,l(x,z)m(x,z)+2\,l(x,z)\right ){{\it dz}}^{2}
\end{equation}
 with arbitrary functions $l(x,z)$ and $m(x,z)$ (\cite{dryuma:dr1}).

   The condition on the curvature tensor
\[
R_{i j k l}=0
\]
for the metric (\ref{dryuma:eq2}) lead to the relations
\[
R_{1313}=\left ({\frac {\partial ^{3}}{\partial
{x}^{3}}}l(x,z)-3\,l(x,z){ \frac {\partial }{\partial
x}}l(x,z)+{\frac {\partial }{\partial z}}l( x,z)\right
){y}+\]\[+\left (\!-\!l(x,z){\frac {\partial ^{2}}{\partial {x}^
{2}}}m(x,z)\!+\!{\frac {\partial ^{2}}{\partial x\partial
z}}m(x,z)\!-\!3\, \left ({\frac {\partial }{\partial
x}}l(x,z)\right ){\frac {\partial } {\partial
x}}m(x,z)\!-\!2\,m(x,z){\frac {\partial ^{2}}{\partial {x}^{2}}}
l(x,z)\!-\!{\frac {\partial ^{2}}{\partial {x}^{2}}}l(x,z)\right
)\!-\!\]\[\left(-m(x,z) {\frac {\partial }{\partial
z}}m(x,z)\!+\!m(x,z)l(x,z){\frac {\partial }{
\partial x}}m(x,z)\!+\!\left ({\frac {\partial }{\partial x}}l(x,z)\right
)m(x,z)\!+\!2\,\left ({\frac {\partial }{\partial x}}l(x,z)\right
)\left ( m(x,z)\right )^{2}\right)/y =0,
\]
and
\[
R_{1323}=\left({\frac {\partial }{\partial x}}l(x,z)-{\frac
{\partial }{\partial z}}m (x,z)+2\,\left ({\frac {\partial
}{\partial x}}l(x,z)\right )m(x,z)+l( x,z){\frac {\partial
}{\partial x}}m(x,z)\right)/y=0,
\]
which are equivalent the system of equations for the functions
$l(x,z)$ and $m(x,z)$
\begin{equation} \label{dryuma:eq3}
{\frac {\partial ^{3}}{\partial {x}^{3}}}l(x,z)-3\,l(x,z){\frac {
\partial }{\partial x}}l(x,z)+{\frac {\partial }{\partial z}}l(x,z)
=0,
\end{equation}
\begin{equation} \label{dryuma:eq4}
{\frac {\partial }{\partial x}}l(x,z)-{\frac {\partial }{\partial
z}}m (x,z)+2\,\left ({\frac {\partial }{\partial x}}l(x,z)\right
)m(x,z)+l( x,z){\frac {\partial }{\partial x}}m(x,z) =0.
\end{equation}

   So the following theorem is valid
\begin{theor}
  There is exists a class of 3-dimensional flat  metrics  defined by the
  solutions of the system of equations (\ref{dryuma:eq3}-\ref{dryuma:eq4}).
\end{theor}

     Remark that the first equation of the system  is the famous
     KdV-equation. Its solutions are very well known and may be used in particular to the
      the description of three orthogonal coordinates systems (\cite{dryuma:dr2}).

    Let us consider some examples.

    1.
\[
l(x,z)=-\frac{x}{3z},\quad m(x,z)=-1/2+{F_1}({\frac
{z}{{x}^{3}}}){x}^{-2}.
\]

    2.
\[
l(x,z)=-4\,\left (\cosh(x-4\,z)\right )^{-2} , \quad
m(x,z)=-\frac{1}{2}
\]
\begin{rem}

     In a most simplest case of 3-dim metrics of zero curvature look as
\[
ds^2={y}^{2}{{\it dx}}^{2}+2\,\left (l(x,z){y}^{2}-1/2\right ){\it
dx}\,{ \it dz}+2\,{\it dy}\,{\it dz}+\]\[+\left (\left
(l(x,z)\right )^{2}{y}^{2}- 2\,\left ({\frac {\partial }{\partial
x}}l(x,z)\right )y+l(x,z)\right ){{\it dz}}^{2}
\]
where the function $l(x,z)$ is solution of classical KdV-equation
\[
{\frac {\partial }{\partial z}}l(x,z)-3\,l(x,z){\frac {
\partial }{\partial x}}l(x,z)+{\frac {\partial ^{3}}{\partial {x}^{3}}
}l(x,z)=0.
\]

   In particular the functions
\[
l(x,z)=-\frac{x}{3z},\quad l(x,z)=-4\,\left (\cosh(x-4\,z)\right
)^{-2},
\]
and
\[
l(x,z)=-24\,{\frac {4\,\cosh(2\,x-8\,z)+\cosh(4\,x-64\,z)+3}{\left
(3\,\cosh( x-28\,z)+\cosh(3\,x-36\,z)\right )^{2}}}
\]
give us the examples of such type of metrics.

     In spite of the fact that the determinant of the metric does not depends from
     the function $l(x,z)$ it is possible to distinguish the properties of the metrics
     with help of the eigenvalue equation for the Laplace operator defined on the 1-form
\[
A(x,y,z)=A_idx^i.
\]

   It has the form
\[
g^{ij}\nabla_i\nabla_j A_k-R^l_k A_l=-\lambda A_k.
\]

    In particular case $l(x,z)=0$ and $A_i(x,y,z)=[h(y),q(y),f(y)]$ this equations
    take the form
\[
-1/4\,{\frac {{\frac {d}{dy}}h(y)-4\,\left ({\frac
{d}{dy}}f(y)\right ){y}^{2}-\left ({\frac
{d^{2}}{d{y}^{2}}}h(y)\right )y+4\,\lambda\,h(y
){y}^{3}}{{y}^{3}}}=0,
\]
\[
-1/4\,{\frac {-6\,h(y)-3\,q(y)+3\,\left ({\frac {d}{dy}}q(y)\right
)y+ 4\,\left ({\frac {d}{dy}}h(y)\right )y-\left ({\frac
{d^{2}}{d{y}^{2}} }q(y)\right
){y}^{2}+4\,f(y){y}^{2}+4\,\lambda\,q(y){y}^{4}}{{y}^{4}}}=0,
\]
\[
-1/4\,{\frac {{\frac {d}{dy}}f(y)-\left ({\frac
{d^{2}}{d{y}^{2}}}f(y) \right
)y+4\,\lambda\,f(y){y}^{3}}{{y}^{3}}}=0.
\]

    The solutions of such type system depends from the eigenvalues $\lambda$ and characterize
    the properties of corresponding flat metric.
\end{rem}

\begin{rem}

   In 3-dimensional geometry the density of  Chern-Simons invariant defined by
\begin{equation} \label{dryuma:eq5}
CS(\Gamma)=\epsilon^{i j k}(\Gamma^p_{i q}\Gamma^q_{k
p;j}+\frac{2}{3}\Gamma^p_{i q}\Gamma^q_{j r}\Gamma^r_{k p})
\end{equation}
has an important role (\cite{dryuma:dr4}).

   For the metric (\ref{dryuma:eq2}) one get the expression
\[
CS(\Gamma)=-{\frac {5\,l(x,z){\frac {\partial }{\partial
x}}l(x,z)-5\,{\frac {
\partial }{\partial z}}l(x,z)}{\sqrt {{y}^{2}}}}-\]\[-{\frac {3\,l(x,z){
\frac {\partial }{\partial x}}m(x,z)-4\,\left ({\frac {\partial }{
\partial x}}l(x,z)\right )m(x,z)-3\,{\frac {\partial }{\partial z}}m(x
,z)-2\,{\frac {\partial }{\partial x}}l(x,z)}{{y}^{2}\sqrt
{{y}^{2}}}}.
\]

     Using this formulae for  the first example one get
\[
CS(\Gamma)={\frac {10}{9}}\,{\frac {x{\it
csgn}(y)}{{z}^{2}y}}-10/3\,{F_1}({ \frac {z}{{x}^{3}}}){\it
csgn}(y){z}^{-1}{x}^{-2}{y}^{-3}.
\]

    In a second  example this quantity is
\[
CS(\Gamma)=160\,{\frac {\cosh(x-4\,z)\left (\sinh(x-4\,z)\right
)^{3}{\it csgn}(y )}{y}}.
\]
\end{rem}
\begin{rem}
 By  analogy the properties of zero curvature metrics can be investigated with help of
 solutions of the MKdF-equation.
\end{rem}
    Our construction of ten dimensional space consists from a few steps.

    The first of them is the creation of six dimensional basic space with
    necessary properties.

    With this aim we use the notion of the Riemann extension of a given three dimensional
    space of zero curvature.

\section{Six dimensional Riemann extensions of the metrics\\[1mm] of zero curvature}

   The Riemann extension of riemannian or nonriemannian spaces can be constructed with the
   help of the Christoffel coefficients $\Gamma^i_{jk}$ of corresponding Riemann space or
   with connection coefficients $\Pi^i_{jk}$ in the case
    of the space of affine connection.

   The metrics of the Riemann extension of any n-dimensional riemannian space looks as
\begin{equation} \label{dryuma:eq6}
^{2n}ds^2=-2\Gamma^i_{jk}dx^jdx^k \psi_i+2dx^id\psi_i,
\end{equation}
where $\psi_i$ are an additional coordinates (\cite{dryuma:dr5}).

    The properties of the spaces equipped with the metric (\ref{dryuma:eq6}) and their
    applications  was  first studied in the works of author ~(\cite{dryuma:dr6}-\cite{dryuma:dr15}).

    In the case of  the metric (\ref{dryuma:eq2}) we get the following expressions for nonzero
    Christoffel coefficients
\[
\Gamma^1_{11}=1/2\,{\frac {2\,l(x,z){y}^{2}-1}{y}} ,\quad
\Gamma^2_{11}=1/4\,{\frac {-4\,{y}^{3}{\frac {\partial }{\partial
x}}l(x,z)+8\,l(x,z ){y}^{2}-1}{y}} ,\quad \Gamma^3_{11}=-y,
\]
\[
\Gamma^1_{12}={y}^{-1},\quad \Gamma^2_{12}=1/2\,{y}^{-1},\quad
\Gamma^1_{13}=1/2\,{\frac {\left (2\,l(x,z){y}^{2}-1\right
)l(x,z)}{y}},\]\[ \Gamma^2_{13}=1/4\,{\frac
{-4\,l(x,z){y}^{3}{\frac {\partial }{\partial x}}l(x,z)+8 \,\left
(l(x,z)\right )^{2}{y}^{2}-l(x,z)-4\,\left ({\frac {\partial ^
{2}}{\partial {x}^{2}}}l(x,z)\right ){y}^{2}+2\,\left ({\frac {
\partial }{\partial x}}l(x,z)\right )y}{y}} ,\]\[
\quad \Gamma^3_{13}=-l(x,z)y ,\quad \Gamma^1_{23}={\frac
{l(x,z)}{y}} ,
 \quad \Gamma^2_{23}=1/2\,{\frac {l(x,z)-2\,\left ({\frac {\partial }{\partial x}}l(x,z)
\right )y}{y}} ,\]\[\Gamma^1_{33}=1/2\,{\frac {2\,\left ({\frac
{\partial }{\partial z}}l(x,z)\right )y- 4\,l(x,z)y{\frac
{\partial }{\partial x}}l(x,z)+2\,{\frac {\partial ^{ 2}}{\partial
{x}^{2}}}l(x,z)+2\,\left (l(x,z)\right )^{3}{y}^{2}- \left
(l(x,z)\right )^{2}}{y}} , \]\[4 y \Gamma^2_{33}=-4\,\left
(l(x,z)\right )^{2}{y}^{3}{\frac {\partial }{\partial x}}l(x
,z)-4\,l(x,z){y}^{2}{\frac {\partial ^{2}}{\partial
{x}^{2}}}l(x,z)+4 \,\left ({\frac {\partial }{\partial
z}}l(x,z)\right )y+2\,{\frac {
\partial ^{2}}{\partial {x}^{2}}}l(x,z)+\]\[+8\,{y}^{2}\left ({\frac {
\partial }{\partial x}}l(x,z)\right )^{2}+8\,\left (l(x,z)\right )^{3}
{y}^{2}-8\,l(x,z)y{\frac {\partial }{\partial x}}l(x,z)-\left
(l(x,z) \right )^{2}-4\,\left ({\frac {\partial ^{2}}{\partial
x\partial z}}l( x,z)\right ){y}^{2},
\]
\[\Gamma^3_{33}=-\left (l(x,z)\right )^{2}y+{\frac {\partial }{\partial
x}}l(x,z).
\]

   Using these expressions  and (\ref{dryuma:eq5}) we get the metric of
    extended six dimensional space with coordinates $(x,y,z,u,v,w)$

\begin{equation} \label{dryuma:eq7}
^{6}ds^2=4y\,\left ({\frac {\partial }{\partial x}}l(x,z)\right
){\it dy} \,{\it dv}+\]\[+y/8\,\left (16\,v{\frac {\partial
^{2}}{\partial x\partial z}}l(x,z)-16\,v\left ({\frac {\partial
^{2}}{\partial {x}^{2}}}l(x,z) \right )l(x,z)+32\,v\left ({\frac
{\partial }{\partial x}}l(x,z) \right )^{2}\right ){{\it
dy}}^{2}+\]\[+\left (-4\,l(x,z){\it dy}+ 2\,{\it dz}-2\,l(x,z){\it
dx}\right ){\it dv}+\]\[+1/8\,\left (\!-\!16\,u{ \frac {\partial
}{\partial z}}l(x,z)\!-\!16\,v{\frac {\partial }{\partial
z}}l(x,z)\!-\!16\,w{\frac {\partial }{\partial
x}}l(x,z)\!+\!32\,u\left ({ \frac {\partial }{\partial
x}}l(x,z)\right )l(x,z)\!-\!16\,v\left ({\frac {\partial
}{\partial x}}l(x,z)\right )l(x,z)\right ){{\it dy}}^{2}+\]\[+4\,
v\left ({\frac {\partial }{\partial x}}l(x,z)\right ){\it
dy}\,{\it dz }+2\,{\it dy}\,{\it dw}-2\,l(x,z){\it dy}\,{\it
du}-4\,v\left ({\frac {\partial }{\partial x}}l(x,z)\right
)l(x,z){\it dx}\,{\it dy}+\]\[+{\frac {1/8\,\left (8\,v{\frac
{\partial ^{2}}{\partial {x}^{2}}}l(x,z)-16\,u {\frac {\partial
^{2}}{\partial {x}^{2}}}l(x,z)\right ){{\it dy}}^{2}+ 4\,v\left
({\frac {\partial ^{2}}{\partial {x}^{2}}}l(x,z)\right ){ \it
dx}\,{\it dy}}{y}}+\]\[+{\frac {2\,v\left ({\frac {\partial
}{\partial x}}l(x,z)\right ){{\it dx}}^{2}+1/8\,\left
(-16\,v{\frac {\partial }{
\partial x}}l(x,z)-64\,u{\frac {\partial }{\partial x}}l(x,z)\right ){
\it dx}\,{\it dy}+\left (1/2\,{\it dy}+{\it dx}\right ){\it
dv}}{{y}^{2}}}+\]\[+{\frac{1/8\, \left (-32\,u{\frac {\partial
}{\partial x}}l(x,z)-12\,v{\frac {
\partial }{\partial x}}l(x,z)\right ){{\it dy}}^{2}+\left ({\it dy}+2
\,{\it dx}\right ){\it du}}{{y}^{2}}}+\]\[+{\frac {1/8\,\left
(16\,w+8\,vl( x,z)+64\,ul(x,z)\right ){\it dx}\,{\it dy}+\left
(1/8\,\left (-16\,u-8 \,v\right ){\it dy}+1/8\,\left
(-32\,u-16\,v\right ){\it dx}\right ){ \it dz}}{{y}^
{3}}}+\]\[+{\frac {1/8\,\left
(-16\,vl(x,z)+16\,w+16\,ul(x,z)\right ){{\it dx}}^{ 2}+1/8\,\left
(4\,w+8\,vl(x,z)+28\,ul(x,z)\right ){{\it dy}}^{2}}{{y}^
{3}}}+\]\[+{\frac {1/8\,\left (-8\,u-4\,v\right ){\it dx}\,{\it
dy}+1/8\, \left (-v-2\,u\right ){{\it dy}}^{2}+1/8\,\left
(-8\,u-4\,v\right ){{ \it dx}}^{2}}{{y}^{5}}} ,
\end{equation}

    Tensor Ricci of the metric (\ref{dryuma:eq7}) has only one component
\begin{equation} \label{dryuma:eq8}
R_{zz}=-{\frac {3\,l(x,z){\frac {\partial }{\partial
x}}l(x,z)-{\frac {
\partial }{\partial z}}l(x,z)-{\frac {\partial ^{3}}{\partial {x}^{3}}
}l(x,z)}{y}}
\end{equation}
and on the solutions of the KdF-equation the space is a flat
thereof that its curvature tensor $^{6}R_{ijkl}$
  is expressed through the Ricci tensor and it is also equal to zero on the solutions
  of the KdF-equation.

 In result after the Riemann extension of three dimensional flat space we get
  six dimensional flat space.

    Next step of our construction is receiving of the Ricci flat
    $^{6}R_{ij}=0$ but not a flat $^{6}R_{i j k l}\neq0$ space.

    With this aim we can insert into the expression for the metric (\ref{dryuma:eq7}) additional
    terms as a result of which can be obtained six dimensional Ricci flat but non a
    flat space.

    Remark that there are a lot possibilities to realize that.

    In a simplest case we item an additional term $$2H(x,z)dxdz$$ with unknown function $H(x,z)$
    into the (\ref{dryuma:eq7})
    and in result we get the metric of perturbed space in form
\begin{equation}\label{dryuma:eq8}
{{\it^{6} ds}}^{2}=\left (2\,{y}^{2}v{\frac {\partial }{\partial
x}}l(x,z) -4\,yvl(x,z)-2\,yul(x,z)+1/2\,{\frac {v}{y}}+{\frac
{u}{y}}+2\,yw \right ){{\it dx}}^{2}+2\,\left (-{\frac
{v}{y}}-2\,{\frac {u}{y}} \right ){\it dx}\,{\it dy}+\]\[+2\,\left
(H(x,z)+1/2\,{\frac {vl(x,z)}{y}}+ 2\,yl(x,z)w\right ){\it
dx}\,{\it dz}+\]\[+\left (2\,yv{\frac {\partial ^{2 }}{\partial
{x}^{2}}}l(x,z)-4\,yv\left (l(x,z)\right )^{2}+2\,{y}^{2}v
l(x,z){\frac {\partial }{\partial x}}l(x,z)-v{\frac {\partial }{
\partial x}}l(x,z)+{\frac {l(x,z)u}{y}}-2\,y\left (l(x,z)\right )^{2}u
\right ){\it dx}\,{\it dz}+\]\[+2\,{\it dx}\,{\it du}+2\,\left
(-{\frac {vl (x,z)}{y}}-2\,{\frac {l(x,z)u}{y}}+2\,v{\frac
{\partial }{\partial x}} l(x,z)\right ){\it dy}\,{\it dz}+2\,{\it
dy}\,{\it dv}+\]\[+\left (1/2\,{ \frac {v\left (l(x,z)\right
)^{2}}{y}}-2\,v{\frac {\partial }{
\partial z}}l(x,z)-4\,yv\left ({\frac {\partial }{\partial x}}l(x,z)
\right )^{2}+2\,yw\left (l(x,z)\right )^{2}-2\,u{\frac {\partial
}{
\partial z}}l(x,z)\right ){d{{z}}}^{2}+\]\[+\left (2\,yvl(x,z){\frac {
\partial ^{2}}{\partial {x}^{2}}}l(x,z)-2\,yu\left (l(x,z)\right )^{3}
\right ){d{{z}}}^{2}+\]\[+\left ({\frac {u\left (l(x,z)\right
)^{2}}{y}}-{ \frac {v{\frac {\partial ^{2}}{\partial
{x}^{2}}}l(x,z)}{y}}-2\,{ \frac {u{\frac {\partial ^{2}}{\partial
{x}^{2}}}l(x,z)}{y}}+2\,yv{ \frac {\partial ^{2}}{\partial
x\partial z}}l(x,z)-4\,yv\left (l(x,z) \right )^{3}\right){{\it
dz}}^{2}+\]\[+\left(2\,{y}^{2}v\left (l(x,z)\right )^{2}{\frac
{\partial }{
\partial x}}l(x,z)+4\,vl(x,z){\frac {\partial }{\partial x}}l(x,z)+4\,
ul(x,z){\frac {\partial }{\partial x}}l(x,z)-2\,w{\frac {\partial
}{
\partial x}}l(x,z)\right ){{\it dz}}^{2}+2\,{\it dz}\,{\it dw}.
\end{equation}

    The space with the metric (\ref{dryuma:eq8}) is a Ricci flat on the solutions of the KdF-equation,
    but its curvature tensor  depends from the
    function $H(x,z)$ and in general $^{6}R_{iklm}\neq0$.

    To take the equations of geodesics of the metric
    (\ref{dryuma:eq8})in the case $l(x,z)=0$.

    They are decomposed into two parts.

    The linear system for the coordinates $(u,v,w)$
\[
{\frac {d^{2}}{d{s}^{2}}}u(s)+2\,\left ({\frac {d}{ds}}x(s)\right
)y{\frac {d}{ds}}w(s)+1/2\,{\frac {\left (\left ({\frac
{d}{ds}}x(s)\right )y-2\,\left ({\frac {d}{ds}}y (s)\right
)y\right ){\frac {d}{ds}}v(s)}{{y}^{2}}}+\]\[+1/2\,{\frac {\left
(2\,\left ({\frac {d}{ds}}x(s)\right )y-4\,\left ({\frac
{d}{ds}}y(s) \right )y\right ){\frac
{d}{ds}}u(s)}{{y}^{2}}}+1/2\,{\frac {\left (2 \,\left ({\frac
{d}{ds}}x(s)\right )^{2}{y}^{2}-4\,\left ({\frac {d}{d
s}}y(s)\right )\left ({\frac {d}{ds}}x(s)\right ){y}^{2}\right
)w(s)}{ {y}^{2}}}+\]\[+1/2\,{\frac {\left (6\,\left ({\frac
{d}{ds}}y(s)\right )^{2 }-3\,\left ({\frac {d}{ds}}y(s)\right
){\frac {d}{ds}}x(s)\right )u(s) }{{y}^{2}}}+1/2\,{\frac {2\,\left
({ \frac {d}{ds}}z(s)\right )^{2}{y}^{2}{\frac {\partial
}{\partial z}}H( x,z)+2\,\left ({\frac {d}{ds}}x(s)\right
)^{2}{y}^{3}H(x,z)}{{y}^{2}}} =0,
\]
\[
{\frac {d^{2}}{d{s}^{2}}}v(s)-2\,{\frac {\left ({\frac
{d}{ds}}x(s)\right ){\frac {d}{ds}}u(s)}{y}} -\left ({\frac
{d}{ds}}x(s)\right )^{2}w(s)-1/4\,{\frac {\left ({ \frac
{d}{ds}}x(s)\right )\left (-24\,{\frac {d}{ds}}y(s)+3\,{\frac {d
}{ds}}x(s)\right )u(s)}{{y}^{2}}}-\]\[-{ \frac {\left ({\frac
{d}{ds}}x(s)\right ){\frac {d}{ds}}v(s)}{y}}=0,
\]
\[
{\frac {d^{2}}{d{s}^{2}}}w(s)-2\,{\frac {\left ({\frac
{d}{ds}}x(s)\right ){\frac {d}{ds}}u(s)}{y}} -\left ({\frac
{d}{ds}}x(s)\right )^{2}w(s)-1/4\,{\frac {\left ({ \frac
{d}{ds}}x(s)\right )\left (-24\,{\frac {d}{ds}}y(s)+3\,{\frac {d
}{ds}}x(s)\right )u(s)}{{y}^{2}}}-\]\[-{ \frac {\left ({\frac
{d}{ds}}x(s)\right ){\frac {d}{ds}}v(s)}{y}}=0.
\]

    And nonlinear system for the coordinates $(x,y,z)$
\[
2\,\left ({\frac {d^{2}}{d{s}^{2}}}x(s)\right )y-\left ({\frac
{d}{ds} }x(s)\right )^{2}+4\,\left ({\frac {d}{ds}}y(s)\right
){\frac {d}{ds}} x(s)=0,
\]
\[
4\,\left ({\frac {d^{2}}{d{s}^{2}}}y(s)\right )y-\left ({\frac
{d}{ds} }x(s)\right )^{2}+4\,\left ({\frac {d}{ds}}y(s)\right
){\frac {d}{ds}} x(s)=0,
\]
\[
{\frac {d^{2}}{d{s}^{2}}}z(s)-y\left ({\frac {d}{ds}}x(s)\right
)^{2}=0.
\]

   In the case $l(x,z)\neq0$ the geodesics of the metric (\ref{dryuma:eq8}) are more complicated.

   Testing with the GRTensorII package show that six-dim spaces with  the metric (\ref{dryuma:eq8}) have
    vanishing scalar curvature invariants and from this point they are same with
    the spaces of gravitational waves of the classical Einstein theory.

\section{Four-dim Schwarzschild space-time on
  six-dim Ricci flat background}

      Here we present the construction of ten dimensional space made up on basis of
       six dimensional Ricii flat space (\ref{dryuma:eq8}) and some four dimensional Einstein space-time.

      In the role of the
      Einstein space-time as a case in point the Schwarzschild space-time will be considered.

    The metric can be choice in the form
\begin{equation}\label{dryuma:eq9}
 ^{10}ds^2=\left (2\,yw+{\frac {u}{y}}+1/2\,{\frac
{v}{y}}+2\,{y}^{2}v{\frac {
\partial }{\partial x}}l(x,z)-2\,yul(x,z)-4\,yvl(x,z)\right ){{\it dx}
}^{2}+2\,\left (-2\,{\frac {u}{y}}-{\frac {v}{y}}\right ){\it
dx}\,{ \it dy}+\]\[+2\,\left (2\,yl(x,z)w-2\,y\left (l(x,z)\right
)^{2}u+{\frac {l (x,z)u}{y}}+2\,yv{\frac {\partial ^{2}}{\partial
{x}^{2}}}l(x,z)+2\,{y }^{2}vl(x,z){\frac {\partial }{\partial
x}}l(x,z)\right ){\it dx}\,{ \it dz}+\]\[+\left (1/2\,{\frac
{vl(x,z)}{y}}-v{\frac {\partial }{\partial x}}l(x,z)-4\,yv\left
(l(x,z)\right )^{2}+H(x,z,r)\right ){\it dx}\,{ \it dz}+2\,{\it
dx}\,{\it du}+\]\[+2\,\left (2\,v{\frac {\partial }{
\partial x}}l(x,z)-2\,{\frac {l(x,z)u}{y}}-{\frac {vl(x,z)}{y}}\right
){\it dy}\,{\it dz}+2\,{\it dy}\,{\it dv}+\]\[+\left (-4\,yv\left
({\frac {
\partial }{\partial x}}l(x,z)\right )^{2}-2\,w{\frac {\partial }{
\partial x}}l(x,z)-4\,yv\left (l(x,z)\right )^{3}+4\,vl(x,z){\frac {
\partial }{\partial x}}l(x,z)\right ){{\it dz}}^{2}+\]\[+\left (2\,{y}^{2}v
\left (l(x,z)\right )^{2}{\frac {\partial }{\partial
x}}l(x,z)+4\,ul(x ,z){\frac {\partial }{\partial
x}}l(x,z)+2\,yvl(x,z){\frac {\partial ^ {2}}{\partial
{x}^{2}}}l(x,z)-{\frac {v{\frac {\partial ^{2}}{
\partial {x}^{2}}}l(x,z)}{y}}\right ){{\it dz}}^{2}+\]\[+\left ({\frac {u
\left (l(x,z)\right )^{2}}{y}}-2\,v{\frac {\partial }{\partial
z}}l(x, z)+2\,yw\left (l(x,z)\right )^{2}-2\,yu\left (l(x,z)\right
)^{3} \right ){{\it dz}}^{2}-\]\[-\left (2\,u{\frac {\partial
}{\partial z}}l(x,z )+2\,{\frac {u{\frac {\partial ^{2}}{\partial
{x}^{2}}}l(x,z)}{y}}-2\, yv{\frac {\partial ^{2}}{\partial
x\partial z}}l(x,z)-1/2\,{\frac {v \left (l(x,z)\right
)^{2}}{y}}\right ){{\it dz}}^{2}+\]\[+2\,{\it dz}\,{ \it dw}+\left
(-{\frac {{c}^{2}M}{r}}+{c}^{2}\right ){{\it dt}}^{2}-{{ \it
dr}}^{2}\left (1-{\frac {M}{r}}\right )^{-1}-{r}^{2}{{\it
d\theta}} ^{2}-{r}^{2}\left (\sin(\theta)\right )^{2}{{\it
d\phi}}^{2}
 \end{equation}
where $H(x,z,r)$ is some arbitrary function.

     The condition $^{10}R_{ik}=0$ on the Ricci tensor of the
     metric (\ref{dryuma:eq9}) lead to the equations
\[
-1/2\,{\frac {-2\,\left ({\frac {\partial }{\partial r}}H(x,z,r)
\right )r-\left ({\frac {\partial ^{2}}{\partial {r}^{2}}}H(x,z,r)
\right ){r}^{2}+\left ({\frac {\partial }{\partial
r}}H(x,z,r)\right ) M+\left ({\frac {\partial ^{2}}{\partial
{r}^{2}}}H(x,z,r)\right )rM}{ {r}^{2}}}=0
\]
and
\[-2\,{\frac {-{\frac {\partial }{\partial z}}l(x,z)+3\,l(x,z){\frac {
\partial }{\partial x}}l(x,z)-{\frac {\partial ^{3}}{\partial {x}^{3}}
}l(x,z)}{y}}=0.
\]

   The solution of the first equation is defined by
\[
H(x,z,r)=F_1(x,z)+F_2(x,z)\left (-\ln (r)+\ln (M-r) \right )
\]
where $F_{i}(x,z)$ are an arbitrary functions and the second
equation is the famous KdF-equation.

   As result the family of the metrics of ten dimensional Einstein spaces with nonzero curvature
   made up on basis of Schwarzschild space and of some six dimensional space  depended  on the solutions
    of KdF-equation has been constructed.

    So the properties of the classical Schwarzschild space and corresponding six
dimensional basic space depending from the solutions of
KdF-equation are interconnected.

  In particular we get new types of trajectories in the
Schwarzschild space-time which are defined by these solutions.

    This statement follows from  the properties of equations of
    geodesics for coordinates $r,x,z$ of the metric (\ref{dryuma:eq9})
\[
{\frac {d^{2}}{d{s}^{2}}}r(s)-1/2\,{\frac {M\left ({\frac
{d}{ds}}r(s) \right )^{2}}{r\left (r-M\right )}}-\left ({\frac
{d}{ds}}\theta(s) \right )^{2}\left (r-M\right )-\left (-r\left
(\cos(\theta)\right )^{2 }+r-M+M\left (\cos(\theta)\right
)^{2}\right )\left ({\frac {d}{ds}} \phi(s)\right
)^{2}+\]\[+1/2\,{\frac {\left (r-M\right ){c}^{2}M\left ({ \frac
{d}{ds}}t(s)\right )^{2}}{{r}^{3}}}+{\frac {\left (r-M\right )
\left ({\frac {\partial }{\partial r}}H(x,z,r)\right )\left
({\frac {d }{ds}}z(s)\right ){\frac {d}{ds}}x(s)}{r}}=0,
\]
\[
{\frac {d^{2}}{d{s}^{2}}}x(s)+1/2\,{\frac {\left (-4\,l(x,z)\left
({ \frac {\partial }{\partial x}}l(x,z)\right )y-\left
(l(x,z)\right )^{2 }+2\,{\frac {\partial ^{2}}{\partial
{x}^{2}}}l(x,z)+2\,{y}^{2}\left ( l(x,z)\right )^{3}+2\,\left
({\frac {\partial }{\partial z}}l(x,z) \right )y\right )\left
({\frac {d}{ds}}z(s)\right )^{2}}{y}}-\]\[-\left (-2 \,{\frac
{l(x,z){\frac {d}{ds}}y(s)}{y}}-1/2\,{\frac {\left (4\,{y}^{2
}\left (l(x,z)\right )^{2}-2\,l(x,z)\right ){\frac
{d}{ds}}x(s)}{y}} \right ){\frac {d}{ds}}z(s)+\]\[+2\,{\frac
{\left ({\frac {d}{ds}}y(s) \right ){\frac
{d}{ds}}x(s)}{y}}+1/2\,{\frac {\left (-1+2\,{y}^{2}l(x, z)\right
)\left ({\frac {d}{ds}}x(s)\right )^{2}}{y}}=0,
\]
\[
{\frac {d^{2}}{d{s}^{2}}}z(s)-\left (-{\frac {\partial }{\partial
x}}l (x,z)+y\left (l(x,z)\right )^{2}\right )\left ({\frac
{d}{ds}}z(s) \right )^{2}-2\,yl(x,z)\left ({\frac
{d}{ds}}z(s)\right ){\frac {d}{ds }}x(s)-y\left ({\frac
{d}{ds}}x(s)\right )^{2}=0,
\]
where the function  $l(x,z)$ is solution of of
    the KdF -equation.

     Remark that full ten dimensional space has nonzero scalar curvature invariants.
    The simplest of them is
$$ C_{i k l m}C^{iklm}=12\,{\frac {{M}^{2}}{{r}^{6}}}
 $$
  where
$C_{i k l m}$ is the Weyl tensor of the metric (\ref{dryuma:eq9}).

   In conclusion it may be sad that suggested method of
   investigation of the properties of classical Einstein spaces
   admits a large generalization from the various points of view.


\begin{thebibliography}
\footnotesize
\bibitem{dryuma:dr1} V.S. Dryuma, Toward a theory of spaces of constant curvature, {Theoretical and Mathematical
Physics}, v.146(1): 2006,  p. 35-45,  {\it ArXiv:
math.DG/0505376}, 18 May 2005, p.1-12.

\bibitem{dryuma:dr2} V.S.Dryuma, Projective propeties of a family operators,{\it IX Vsesoyuznaya Geometricheskaya
 konferenciya}, Kishinev, 20-22 Sentyabrya,1988 g., Theses of communications, pp.104-105 (in Russian).

 \bibitem{dryuma:dr3} V.S.Dryuma,
 Three dimensional exactly integrable system of nonlinear equations and its applications,
 {\it Matematicheskie issledovaniya}, Kishinev, Stiintsa, 1992, v.124, pp.56-68 (in Russian).

\bibitem{dryuma:dr4} Jackiw R., A Pure Cotton Kink in a Funny Place,
 {\it ArXiv: math-ph/0403044}, v.2, 21 July, 2004.

\bibitem{dryuma:dr5} Paterson E.M. and Walker A.G.,Riemann extensions,{\it Quart.J.Math.Oxford},
 1952, V.3,19--28.

\bibitem{dryuma:dr6} The Riemann Extension in theory of
differential equations and their
  applications,{\it Matematicheskaya fizika, analiz, geometriya}, 2003,v.10, No.3,1--19.


 \bibitem{dryuma:dr7} Dryuma V.,Applications of Riemannian and Einstein-Weyl Geometry in
 the theory of second order ordinary differential equations,{\it Theoretical and
 Mathematical Physics}, 2001, V.128, N 1, 845--855.


 \bibitem{dryuma:dr8} Dryuma V., The Riemann and Einstein-Weyl geometries in the theory of ODE's
 , their applications and all that,
{New Trends in Integrability and Partial Solvability ,115-156,
(eds.A.B.Shabat et al.)  Kluwer Academic Publishers.}, { ArXiv:
nlin: SI/0303023, 11 March, 2003, 1--37}.


\bibitem{dryuma:dr9} Dryuma V.,The Riemann Extension of the Schwarzschild space-time,
 {\it  Buletinul A.\c S. M. Matematica, Nr.3(43)}, 2003, p.92-103.

 \bibitem {dryuma:dr10} Dryuma V. On geodesical
extension of the Schwarzshild space-time,  {\it Proceedings of
Institute of Mathematics of NAS of Ukraine}, Kiev, v.50, 2004,
p.1294-1300.

\bibitem {dryuma:dr11} Dryuma V. On the Riemann
extension of the space-time with
 vanishing curvature invariants, {\it http://www.math.kth.se/4ecm/abstract/14.27.pdf},  Sweden,
Stockholm, June 27-July 2, ECM'2004.

\bibitem{dryuma:dr12} Dryuma V.,The Riemann Extension of the Minkowsky space-time
 in rotating coordinate system, { \it ArXiv: gr-qc/0505122}, v1, 24 May 2005, p.1-10.

\bibitem {dryuma:dr13} Dryuma V., Riemann extensions in theory of
the first order systems of differential equations, {\it
ArXiv:math.DG/0510526},  25 October 2005, p.1-21.

\bibitem {dryuma:dr14}  Dryuma., From the Number theory to the
Riemann geometry, {\it Scientific Papers International Conference
"Differential Equations and Computer Algebra Systems (DE and
CAS'2005)"}, Minsk, 2005,  BGPU,  Part 1, 14-15.

\bibitem{dryuma:dr15} Dryuma V.,On the Riemann extensions of the G\"odel space-time metric,
 {\it  Buletinul A.\c S. M. Matematica, Nr.3(49)}, 2005, p.1-20.  {\it
ArXiv:gr-qc/0511165},  30 Nov 2005, p.1-12.
 \end{thebibliography}
 \end{document}